\documentclass[12pt]{article}
\usepackage{amsmath,amssymb,color}

\usepackage[text={17cm,22.5cm}]{geometry}

\DeclareMathOperator\atanh{atanh}
\def\red#1{{\color{red} #1}}

\let\OLDthebibliography\thebibliography
\renewcommand\thebibliography[1]{
  \OLDthebibliography{#1}
  \setlength{\parskip}{3pt}
  \setlength{\itemsep}{3pt plus 0.2ex}  
}

\begin{document}

\def\prg#1{\par\medskip\noindent{\bf #1}}  \def\ra{\rightarrow}
\newcounter{nbr}
\def\note#1{\bitem\vspace{-5pt}\addtocounter{nbr}{1}
            \item{} #1\vspace{-5pt}
            \eitem}
\def\lra{\leftrightarrow}           \def\Ra{\Rightarrow}
\def\nin{\noindent}                 \def\pd{\partial}
\def\dis{\displaystyle}             \def\Lra{{\Leftrightarrow}}
\def\tgr{{GR$_{||}$}}               \def\tmgl{\hbox{TMG$_\Lambda$}}
\def\vsm{\vspace{-9pt}}             \def\vsmb{\vspace{-5pt}}
\def\cs{{\scriptstyle\rm CS}}       \def\ads3{{\rm AdS$_3$}}
\def\Leff{\hbox{$\mit\L_{\hspace{.6pt}\rm eff}\,$}}
\def\bull{\raise.25ex\hbox{\vrule height.8ex width.8ex}}
\def\Lie{{\cal L}\hspace{-.7em}\raise.25ex\hbox{--}\hspace{.2em}}
\def\sS{\hspace{2pt}S\hspace{-0.83em}\diagup}
\def\ric{{Ric}}
\def\nb{ \marginpar{\bf\Large ?} }  \def\hd{{^\star}}
\def\dis{\displaystyle}             \def\mb#1{\hbox{{\boldmath $#1$}}}
\def\ul#1{\underline{#1}}           \def\ub#1{\underbrace{#1}}
\def\phb{\phantom{\Big|}}
\def\chm{\checkmark}                \def\chmr{\red{\chm}}
\def\ir#1{\,{}^{#1}\hspace{-1.2pt}}
\def\irr#1{\,{}^{(#1)}\hspace{-1.2pt}}  \def\em{\text{em}}  \def\gr{\text{GR}}

\def\hook{\hbox{\vrule height0pt width4pt depth0.3pt
\vrule height7pt width0.3pt depth0.3pt
\vrule height0pt width2pt depth0pt}\hspace{0.8pt}}
\def\inn{\hook}
\def\first{\rm (1ST)}  \def\second{\hspace{-1cm}\rm (2ND)}

\def\G{\Gamma}        \def\S{\Sigma}        \def\L{{\mit\Lambda}}
\def\D{\Delta}        \def\Th{\Theta}
\def\a{\alpha}        \def\b{\beta}         \def\g{\gamma}
\def\d{\delta}        \def\m{\mu}           \def\n{\nu}
\def\th{\theta}       \def\k{\kappa}        \def\l{\lambda}
\def\vphi{\varphi}    \def\ve{\varepsilon}  \def\vth{{\vartheta}}
\def\p{\pi}           \def\chk{\checkmark}
\def\r{\rho}          \def\Om{\Omega}       \def\om{\omega}
\def\s{\sigma}        \def\t{\tau}          \def\eps{\epsilon}
\def\nab{\nabla}      \def\btz{{\rm BTZ}}   \def\heps{{\hat\eps}}

\def\bT{\bar{T}}      \def\hH{\widehat{H}}  \def\hE{\widehat{E}}
\def\tG{{\tilde G}}   \def\cF{{\cal F}}    \def\cA{{\cal A}}
\def\cL{{\cal L}}     \def\cM{{\cal M }}   \def\cE{{\cal E}}
\def\cH{{\cal H}}     \def\hcH{\hat{\cH}}  \def\cT{{\cal T}}
\def\hA{\hat{A}}      \def\hB{\hat{B}}     \def\hK{\hat{K}}
\def\cK{{\cal K}}     \def\hcK{\hat{\cK}}  \def\cT{{\cal T}}
\def\cO{{\cal O}}     \def\hcO{\hat{\cal O}} \def\cV{{\cal V}}
\def\tom{{\tilde\omega}}  \def\cE{{\cal E}} \def\bH{\bar{H}}
\def\cR{{\cal R}}    \def\hR{{\hat R}{}}   \def\hL{{\hat\L}}
\def\tb{{\tilde b}}  \def\tA{{\tilde A}}   \def\hom{{\hat\om}}
\def\tL{{\tilde L}}  \def\tR{{\tilde R}}   \def\tcL{{\tilde\cL}}
\def\he{{\hat e}}    \def\hom{{\hat\om}}   \def\hth{\hat\theta}
\def\hxi{\hat\xi}    \def\hg{\hat g}       \def\hb{{\hat b}}
\def\tH{{\tilde H}}  \def\tK{{\tilde K}}   \def\ha{{\bar a}}
\def\hb{{\bar b}}    \def\bR{\bar{R}}      \def\bF{{\bar F}}
\def\haa{{\bar\a}}   \def\hbb{{\bar\b}}    \def\hgg{{\bar\g}}
\def\tPhi{{\tilde\Phi}} \def\barb{{\bar b}} \def\tPsi{{\tilde\Psi}}
\def\bK{{\bar K}}    \def\bk{{\bar k}}     \def\orth{{\perp}}
\def\bi{{\bar\imath}} \def\bj{{\bar\jmath}} \def \bk{{\bar k}}
\def\bm{{\bar m}}     \def\bn{{\bar n}}    \def\bl{{\bar l}}
\let\Pi\varPi         \def\chH{\check{H}}  \def\bB{{\bar B}}
\let\eR\varOmega    \def\hpi{{\hat\pi}}    \def\hPi{{\hat\Pi}}
\def\cN{{\cal N}}    \def\cB{\cal B}       \def\heps{{\hat\epsilon}}
\def\hN{{\hat N}}    \def\bb{{\bar b}}     \def\fL{L^\text{1st}}
\def\T{\mathbb{T}}   \def\emph#1{{\it #1}}

\vfuzz=2pt 
\def\nn{\nonumber}
\def\be{\begin{equation}}             \def\ee{\end{equation}}
\def\ba#1{\begin{array}{#1}}          \def\ea{\end{array}}
\def\bea{\begin{eqnarray} }           \def\eea{\end{eqnarray} }
\def\beann{\begin{eqnarray*} }        \def\eeann{\end{eqnarray*} }
\def\beal{\begin{eqalign}}            \def\eeal{\end{eqalign}}
\def\lab#1{\label{eq:#1}}             \def\eq#1{(\ref{eq:#1})}
\def\balign{\begin{align}}            \def\ealign{\end{align}}
\def\bsubeq{\begin{subequations}}     \def\esubeq{\end{subequations}}

\def\bitem{\begin{itemize}\vspace{-1pt} \setlength\itemsep{-4.5pt} }
  \def\eitem{\end{itemize}\vspace{-1pt} }
\renewcommand{\theequation}{\thesection.\arabic{equation}}
\def\aff#1{\vspace{-12pt}{\normalsize #1}}

\title{On thermodynamics of charged black hole with scalar hair}

\author{M. Blagojevi\'c and B. Cvetkovi\'c\footnote{
        Email addresses: \texttt{mb@ipb.ac.rs, cbranislav@ipb.ac.rs}} \\
\aff{Institute of Physics, University of Belgrade,
                           Pregrevica 118, 11080 Belgrade, Serbia} }
\date{}
\maketitle

\begin{abstract}
It is shown that energy, entropy and the first law of the Martinez-Troncoso black hole with electric charge and scalar hair (2006) can be consistently described in a general Hamiltonian approach to black hole thermodynamics.
\end{abstract}


\section{Introduction}\label{sec1}

The concept of black hole entropy introduced in the 1970s \cite{bekenstein-1973}
had a long-term influence on our understanding of the gravitational dynamics. During several decades, this influence has been limited to Riemannian black holes, obtained as solutions of Einstein's general relativity (GR) or higher derivative gravity \cite{wald-1993, jacobson-1994}. However, in the early 1960s, there appeared a modern, gauge-field-theoretic theory of gravity, known as the Poincar\'e gauge theory (PG) \cite{kibbe-1961, fh-1980, mb.fh-2013,mb-2002,yo-2006}. In PG, spacetime is characterized by a Riemann-Cartan geometry in which both the torsion and the curvature influence the gravitational dynamics. As a consequence, PG offers new possibilities for exploring the interplay between dynamics, geometry and black hole entropy.

Through the years, many well-known black hole solutions of GR have been successfully generalized to the PG solutions with torsion \cite{mb.fh-2013}, but a consistent analysis of their thermodynamic properties has long been missing. However, since recently, there exists a rather general Hamiltonian approach to black hole entropy \cite{mb.bc-2019} which extends the concept of entropy from Riemannian to Riemann-Cartan spacetimes. The approach offers an efficient description of entropy and the first law of black hole thermodynamics not only in PG, but also in its Riemannian (vanishing torsion) or teleparallel (vanishing curvature) subsectors  \cite{mb.bc-2022,mb.bc-2020}.

The physics of black holes in the 1960s supported the idea that ``a black hole has no hair", see Ruffini and Wheeler \cite{wheeler-1971}. According to this no-hair conjecture, black holes cannot have any other charge except for mass, angular momentum and electric/magnetic charge \cite{bekenstein-1997}. Since the 1990s, attempts to clarify the range of validity of this conjecture led to discovering a plethora of new, ``hairy" black holes as counterexamples to the no-hair conjecture. Convincing results of our Hamiltonian approach in interpreting entropy as the canonical charge on horizon motivated us to examine its further extension to \emph{hairy black holes}.

Martinez et al. \cite{martinez-2004} reported an exact four-dimensional black hole solution of GR with scalar hair, which is asymptotically locally AdS. Relying on our Hamiltonian approach, we found a simple and consistent description of its energy and entropy \cite{mb.bc-2023}. In the present paper, we focus our attention on the subsequent work of Martinez and Troncoso (MT) \cite{martinez-2006} representing a natural generalization of \cite{martinez-2004}, with Maxwell field as an additional part of the matter sector.

The paper is organized as follows. In section \ref{sec2}, we present a PG-inspired tetrad formulation of the MT black hole as a Riemannian solution of GR. In section \ref{sec3}, we introduce general boundary terms at infinity and horizon, use them to calculate energy and entropy as the corresponding canonical charges, and verify the first law. In section \ref{sec4}, we analyze boundary terms for the MT black hole considered as a solution of teleparallel gravity. Section \ref{sec5} is devoted to concluding remarks, and Appendix contains some technical details.

Our conventions are the same as in Ref. \cite{mb.bc-2023}. Latin indices $(i, j,\dots)$ are the local Lorentz indices, greek indices $(\mu,\nu,\dots)$ are the coordinate indices, and both run over $0,1,2,3$; the orthonormal coframe (tetrad) is $\vth^i=\vth^i{}_\m dx^\m$ (1-form), $\vth:=\det(\vth^i{}_\m)$, the dual basis (frame) is $e_i= e_j{}^\m\pd_\m$, and $\om^{ij}=\om^{ij}{}_\m dx^\m$ is the metric compatible connection (1-form); the metric components in the local Lorentz basis are $\eta_{ij}=(1,-1,-1,-1)$, and the totally antisymmetric symbol $\ve_{ijmn}$ is normalized by $\ve_{0123}=1$; the Hodge dual of a form $\a$ is denoted by $\hd\a$, and the wedge product of forms is implicitly understood

\section{The MT black hole in the tetrad formalism}\label{sec2}
\setcounter{equation}{0}

\subsection{Dynamics}

In our study of the MT black hole entropy, we use the general PG formalism  \cite{fh-1980, mb.fh-2013,mb-2002,yo-2006}, where the tetrad field $\vth^i$ and the metric compatible spin connection $\om^{ij}$ are apriori \emph{independent} dynamical variables, interpreted as the gauge potentials associated to the local Poincar\'e symmetry. The corresponding field strengths, the torsion $T^i:=d\vth^i+\om^i{}_k\vth^k$ and the curvature $R^{ij}=d\om^{ij}+\om^i{}_k\om^{kj}$, characterize the Riemann-Cartan geometry of spacetime.

Consider a system of the gravitational field coupled to matter consisting of the scalar and electromagnetic field, described by the Lagrangian
\be
L=L_G+L_\phi+L_\em\,,                                           \lab{2.1}
\ee
where
\bsubeq
\bea
&&L_G:=-a_0\hd R\equiv -a_0\hd(\vth_i\vth_j)R^{ij}\,,             \nn\\
&&L_\phi:=\frac{1}{2}d\phi\hd d\phi+\hd V(\phi)\,, \qquad
  L_\em:=-\frac{1}{16\pi}F\,\hd F\,.
\eea
Here, $a_0=1/16\pi G$, the gravitational part is \emph{linear in curvature}\footnote{In the framework of PG, the Lagrangian $L_G$ defines the Einstein--Cartan theory of gravity \cite{mb.fh-2013}.},  the potential $V(\phi)$ describes a nonlinear self-interaction of the scalar field,
\be
 V(\phi)=\frac{k}{2\ell^2}\cosh^4\Big(\frac{\phi}{\sqrt{k}}\Big)\,,
\ee
\esubeq
where $k=12a_0$ and $F=dA$ is the electromagnetic field strength.

Since $L_G$ is linear in curvature, its variation with respect to $\om^{ij}$ yields the condition of vanishing torsion, whereupon $\om^{ij}$ becomes a Riemannian connection \cite{mb.bc-2020,mb.bc-2023}. Thus, we end up with a Riemannian spacetime, the subcase of PG with $T^i=0$, where the tetrad field $\vth^i$ is the only independent gravitational variable. A complementary description of the MT black hole, based on the teleparallel geometry with $R^{ij}=0$, will be discussed in section \ref{sec4}.

After introducing the matter covariant momenta
\be
H_\phi:=\frac{\pd L_\phi}{\pd d\phi}=\hd d\phi\,,\qquad
H_{\em}:=\frac{L_\em}{\pd F}=-\frac{1}{4\pi}\hd F\,,
\ee
the variation of the Lagrangian \eq{2.1} with respect to $\phi,A$ and $\vth^i$ yields the field equations
\bsubeq\lab{2.4}
\bea
&&\cE_\phi:=-dH_\phi+\pd_\phi\hd V=0\,,          \lab{2.4a}\\
&&\cE_\em:=-d H_\em=0\,,                         \lab{2.4b}\\
&&\cE_i:=E_i+\t_i+\cT_i=0\,,                     \lab{2.4c}
\eea
\esubeq
where
\be
 E_i:=\frac{\pd L_G}{\pd\vth^i}\,,\qquad
\t_i=:\frac{\pd L_\phi}{\pd\vth^i}\,,\qquad \cT_i:=\frac{\pd L_\em}{\pd\vth^i}\,,
\ee
are the gravitational and matter energy-momentum currents (3-forms), respectively.

\subsection{Geometry}

The metric of the MT spacetime is static and spherically symmetric,
\bsubeq\lab{2.6}
\be
ds^2=f^2 dt^2-\frac{dr^2}{g^2}-r^2(d\r^2+\sinh^2\r\, d\vphi^2)\,,   \lab{2.6a}
\ee
where
\bea
&&f^2=N^2C^{-1}\,,\qquad g^2=N^2 C^2\,,                 \nn\\
&&N^2:=\frac{r^2}{\ell^2}-1+G\frac{q^2}{\ell^2}\,,\qquad
  C:=1+G\frac{q^2}{r^2}\,.
\eea
\esubeq
The only nontrivial parameter of the solution is $q$; it is proportional to the electric charge (subsection \ref{sub33}). The horizon radius is defined as the positive root of $N^2=0$,
\be
r_+=\sqrt{\ell^2-Gq^2}\,.
\ee
The tetrad field associated to the MT metric \eq{2.6a} is chosen in the diagonal form
\be
\vth^0:=f dt\,,\qquad \vth^1:=\frac{dr}{g}\,,\qquad
\vth^2=rd\r\,,\qquad \vth^3:=r\sinh\r\,d\vphi\,.                    \lab{2.8}
\ee
The horizon area is given by
\be
\cA_H=\int_{S_H}\vth^2\vth^3=r_+^2\s\,,                             \lab{2.9}
\ee
where $\s$ is the horizon area normalized to $r_+=1$.
The black hole temperature, which is determined by surface gravity, does not depend on the electric charge parameter $q$,
\be
\k:=g\pd_r f\big|_{r_+}=\frac{1}{\ell}\quad\Ra\quad T:=\frac{1}{2\pi\ell}\,,
\ee
see also \cite{vertogradov-2023}. The Riemannian spin connection reads (with $c=2,3$)
\be
\om^{01}=-\frac{g}{f}f'\vth^0\,,\qquad \om^{1c}=\frac{g}{r}\vth^c\,,
  \qquad\om^{23}=\frac{\cosh\th}{r\sinh\th}\vth^3\,.             \lab{2.11}
\ee

One can show that the tetrad \eq{2.8} combined with the matter fields
\be
\phi:=\sqrt{k}\atanh\left( \sqrt{\frac{Gq^2}{r^2+Gq^2}}\,\right)\,,\qquad
A:=-\frac{q}{\sqrt{r^2+Gq^2}}\frac{\vth^0}{f}\,,                 \lab{2.12}
\ee
solves the field equations  \eq{2.4} provided
\be
k=12a_0=\frac{3}{4\pi G}\,.
\ee
For technical details, see Appendix A.

The nonvanishing value of $V$ at $q=0$ plays the role of an effective cosmological constant, $V(0)=6a_0/\ell^2=:-2\L_\text{eff}$,  associated to an \emph{asymptotically AdS background}. Indeed, for $q=0$, the metric functions take the form $f^2=g^2=r^2/\ell^2-1$. As a consequence, the curvature scalar $R$ is not maximally symmetric but contains additional $O(r^{-1})$ terms, which prevents the MT black hole to be a solution of the quadratic curvature gravity; for details, see Ref. \cite{mb.bc-2023}.

\section{Boundary terms}\label{sec3}
\setcounter{equation}{0}

Thermodynamic variables of the MT black hole are defined by the gravitational and matter contributions to the boundary integral $\G:=\G_\infty-\G_H$, determined by the following variational equations:
\bsubeq\lab{3.1}
\bea
\d\G_\infty&=&\oint_{S_\infty}\d B(\xi)\,,\qquad
       \d\G_H=\oint_{S_H} \d B(\xi)\,,                               \\
\d B(\xi)&:=&(\xi\inn\vth^{i})\d H_i+\d\vth^i(\xi\inn H_i)
   +\frac{1}{2}(\xi\inn\om^{ij})\d H_{ij}
   +\frac{1}{2}\d\om^{ij}(\xi\inn\d H_{ij})                          \nn\\
&&-\d\phi(\xi\inn H_\phi)+(\xi\inn A)\d H_\em+\d A(\xi\inn H_\em)\,.\lab{3.1b}
\eea
\esubeq
Here, $\xi$ is the Killing vector for local translations (for static black holes, $\xi=\pd_t$), and $(S_\infty,S_H)$ are components of the boundary of the spacial section $\S$ of spacetime, located at infinity and horizon, respectively. When the boundary integrals $(\G_\infty,\G_H)$ are finite, they define \emph{energy and entropy} as the canonical charges at infinity and horizon, respectively. The upper line in \eq{3.1b} describes the gravitational, and the lower one the matter (scalar and Maxwell) contributions to $\d B$. The variation $\d$ is assumed to satisfy the following general requirements:
\bitem
\item[(r1)] On $S_\infty$, $\d$ acts on the parameters of the solution, but not on the background configuration.
\item[(r2)] On $S_H$, surface gravity is constant, $\d\k=0$.
\eitem

The derivation of the formulas \eq{3.1} is based on the regularity of the corresponding canonical gauge generator \cite{regge-1974}, which is ensured by the condition
\be
\d\G\equiv\d\G_\infty-\d\G_H=0\,.                                   \lab{3.2}
\ee
This formula relates energy, entropy and Maxwell charge in a way that represents the \emph{first law of black hole thermodynamics}. For more details, see Refs. \cite{mb.bc-2019,mb.bc-2022,mb.bc-2023}.

\subsection{The gravitational contribution}

In Riemannian spacetime, where $H_i=0$, the nontrivial covariant momenta are
$H_{ij}=-2a_0\hd(\vth_i\vth_j)$, and the gravitational contribution to the  boundary term reads
\bea
\d B_G&=&\om^{01}{}_t\d H_{01}+\d\om^{12}H_{12t}+\d\om^{13}H_{13t} \lab{3.3}\\
 &=&2a_0f'g\d(\vth^2\vth^3)-2a_0\d\Big(\frac{g}{r}\vth^2\Big)f\vth^3
   +2a_0\d\Big(\frac{g}{r}\vth^3\Big)f\vth^2                        \nn\\
 &=&2a_0f'g\d(r^2\s)-4a_0\big(\d g\big)fr\s\,.                      \nn
\eea
where we use $\om^{ij}{}_t\equiv\xi\inn\om^{ij}$ and $H_{ijt}\equiv\xi\inn H_{ij}$.

Energy is obtained by calculated $\d B_G$ at infinity. Since $r^2\big|_{\infty}$ does not depend on the black hole parameter $q$, the first term vanishes, so that
\bsubeq\lab{3.4}
\be
(\d\G_G)_\infty=-12a_0\frac{q\d q}{\ell^2}r\s+O_1\,.             \lab{3.4a}
\ee
As we shall see, this divergence will be cancelled by the corresponding scalar field contribution.

Entropy is determined by calculating $\d\G_G$ at horizon:
\be
(\d\G_G)_H=2a_0(gf')_H\d r_+^2\s=2a_0\k\d r_+^2\s=T\d S\,,\qquad
     S:=\frac{r_+^2\s}{4G}\,.                                     \lab{3.4b}
\ee
\esubeq

\subsection{The scalar field contribution}

Now, consider the first term in the lower line of \eq{3.1b}:
\be
\d B_\phi=-\d\phi\,(\xi\inn\hd d\phi)=-\d\phi\,\phi'gf\vth^2\vth^3\,.
\ee

The calculation of energy yields
\be
(\d\G_\phi)_\infty=-\d_q\phi\,\phi'gfr^2\s
                     =k\frac{q\d q}{\ell^2}r\s +O_1\,.
\ee
Hence, for $k=12a_0$, divergent contributions to energy stemming from gravity and the scalar matter cancel each other,
\be
(\d\G_G)_\infty +(\d\G_\phi)_\infty=0\,.                        \lab{3.7}
\ee

On the other hand, $g(r_+)=0$ implies
\be
(\d\G_\phi)_H=0\,.
\ee

\subsection{The Maxwell contribution}\label{sub33}

Using the electromagnetic potential $A$ defined in Eq. \eq{2.12}, one can calculate the corresponding covariant momentum
\be
H_\em=-\frac{1}{4\pi}\hd F =\frac{1}{4\pi}\frac{q}{r^2}\vth^2\vth^3\,,\lab{3.9}
\ee
and obtain the asymptotic electric charge:
\be
Q=\int_{S_\infty}H_\em=\frac{q\s}{4\pi}\,.~
\ee

Now, using the last two boundary terms in \eq{3.1b}, one finds that the electromagnetic contribution to  energy vanishes,
\be
(\d\G_\em)_\infty=\int_{S_\infty} A_t\d H_\em=0\,.                  \lab{3.11}
\ee
where we used $A_t=O_1$. Combining this result with \eq{3.7}, one concludes that the complete energy of the MT black hole vanishes.

The electromagnetic contribution at horizon takes the standard form,
\be
(\d\G_\em)_H=\int_{S_H} A_t\d H_\em
            =\frac{q}{\ell}\frac{\d(q\s)}{4\pi}=\Phi\d Q\,,
\ee
where $\Phi$ is the electromagnetic potential
\be
\Phi:=A_t\Big|^\infty_{r_+}=\frac{q}{\ell}\,.
\ee

\subsection{The first law}

The form of the boundary terms at infinity implies that energy (mass) of the MT black hole vanishes:
\bsubeq
\be
\d E=\d\G_\infty=0\,.
\ee
Similarly, the sum of the boundary terms at horizon also vanishes:
\be
\d\G_H=T\d S+\Phi\d Q=0\,.
\ee
\esubeq
Hence, the first law \eq{3.2} takes the form
\be
\d\G_\infty=\d\G_H\quad\Lra\quad 0=T\d S+\Phi\d Q\,.                \lab{3.15}
\ee

Since the black hole temperature does not depend on the parameter $q$, one can combine the relation $T\d S=\d (TS)$ with $\Phi\d Q=Q\d\Phi$ to rewrite the first law as
\be
\d\left(TS+\frac 12\Phi Q\right)=0\,.
\ee
This relation can be obtained from a hairy deformation of the \emph{Smarr formula} \cite{smarr-1973}
\be
TS+\frac 12\Phi Q-\frac{\s l}{8\pi G}=0\,,                          \lab{3.17}
\ee
where the third term is an extra contribution, independent of the solution parameter $q$.

\section{The MT black hole in teleparallel gravity}\label{sec4}
\setcounter{equation}{0}

In the framework of PG, the teleparallel theory of gravity (TG) is defined by the condition of vanishing curvature \cite{mb.fh-2013}. The TG dynamics is naturally defined by a Lagrangian which is quadratic in torsion,
\be
L_T:=T^i\hd(a_1\irr{1}T^i+ a_2\irr{2}T^i+ a_3\irr{3}T^i)\,,
\ee
where $\irr{n}T^i$ are irreducible components of the torsion.
From the physical point of view, of particular importance is a \emph{one-parameter family} of TG Lagrangians, defined by
\be
(a_1,a_2,a_3)=a_0\times(1,-2,-1/2+\g)\,,
\ee
which is empirically indistinguishable from GR.

By adopting the relation $\om^{ij}=0$ as a gauge fixing condition for local Lorentz symmetry,  torsion takes the simplified form $T^i=d\vth^i$.
To examine thermodynamic properties of the MT solution, we use the  tetrad \eq{2.8} to obtain
\bea
&&T^0=-N'C\vth^0\vth^1\,,\qquad T^2:=\frac{NC}{r}\vth^1\vth^2\,,\nn\\
&&T^3=\frac{1}{r}\big(\coth\th\,\vth^2\vth^3+NC\,\vth^1\vth^3\big)\,.
\eea
Since $\irr{3}T^i=0$, the torsion covariant momentum  $H_i=2a_0\hd\Big(\irr{1}T^i-2\irr{2}T^i\Big)$ does not depend on the parameter $\g$:
\bea
&&H_0=\frac{2a_0}{r}\big(\coth\th\,\vth^1\vth^3
                                -2NC\,\vth^2\vth^3\big)\,,   \nn\\
&&H_1=-\frac{2a_0}{r}\coth\th\,\vth^0\vth^3\,,                \nn\\
&&H_2=\frac{2a_0C}{r}(Nr)'\vth^0\vth^3\,,\nn\\
&&H_3=-\frac{2a_0C}{r}(Nr)'\vth^0\vth^2\,.
\eea

In TG, the general boundary term \eq{3.1b} is reduced to
\be
\d B(\xi)=(\xi\inn\vth^{i})\d H_i+\d\vth^i(\xi\inn H_i)
                                   ~+~\text{matter terms}\,.
\ee
Since the matter part remains the same as in GR, all we need to calculate is the gravitational contribution, the nonvanishing part of which reads
\be
\d B_G=\vth^0{}_t\d H_0+\d\vth^2H_{2t}+\d\vth^3H_{3t}\,.       \lab{4.6}
\ee

Explicit calculation gives the following gravitational boundary terms at infinity and horizon:
\bsubeq\lab{4.7}
\bea
&&\left(\d \G_G\right)_\infty=\int_{S_\infty}\d B_G
                             =-12a_0\s\frac{r}{\ell^2}q\d q+\cO_1\,, \\
&&(\d\G_G)_H=\int_{S_H}\d B_G=2a_0\k\s\d r_+^2=T\d S\,.
\eea
\esubeq
Thus, the final values of the gravitational boundary terms
in GR and in teleparallel gravity, presented respectively in Eqs, \eq{3.4} and \eq{4.7}, coincide. Hence, energy, entropy and the first law of the MT black hole in the one-parameter TG coincide with the corresponding GR results.

\section{Conclusions}\label{sec5}
\setcounter{equation}{0}

In this paper, we used the general Hamiltonian approach proposed in Ref. \cite{mb.bc-2019} to study energy, entropy and the first law of the MT black hole in two complementary geometric settings.

First, we showed that our canonical approach, originally designed for black holes in PG, can be successfully applied to the electrically charged black hole with scalar hair, found by Martinez and Troncoso \cite{martinez-2006} as a \emph{Riemannian} solution of GR.

For vanishing energy and constant temperature, the first law \eq{3.15} is associated to a hairy deformation of the Smarr formula \eq{3.17}.

The MT solution is reinterpreted as an exact solution of the \emph{teleparallel gravity}. Although the original analytic expressions \eq{3.3} and \eq{4.6} for the gravitational boundary terms in GR and TG are different, their final values coincide.

We expect that the present analysis can be consistently extended to other hairy black holes.

\section*{Acknowledgements}

This work was partially supported by the Ministry of Education, Science and Technological development of the Republic of Serbia.

\appendix
\section{Gravitational field equation \eq{2.4c}}\label{appA}
\setcounter{equation}{0}

It is not difficult to verify the validity of the matter field equations \eq{2.4a} and \eq{2.4b}. To prove the gravitational field equation \eq{2.4c}, we find it convenient to express all the energy-momentum currents in terms of the corresponding energy-momentum tensors.

For the scalar field, the energy-momentum current $\t_i$ defines the corresponding energy-momentum tensor $\t^k{}_i$ by
\bea
&&\t_i:=h_i\inn L_\phi-(h_i\inn d\phi)H_\phi=-\heps_k\t^k{}_i\,,        \nn\\ &&\t^k{}_i:= \pd^k\phi\pd_i\phi-\d^k_i\cL_\phi\,,\qquad\cL_\phi:=-\hd L_\phi\,.
\eea
An analogous procedure in the electromagnetic sector yields
\bea
&&\cT_i:=h_i\inn L_\em-(h_i\inn F)H_\em=-\heps_k\cT^k{}_i\,,            \nn\\
&&\cT^k{}_i:=-\frac{1}{4\pi}\left( F^{km}F_{im}
                           -\frac{1}{4}\d^k_i F^{mn}F_{mn}\right)\,,
\eea
and finally, the gravitational energy-momentum current is defined as
\bea
&&E_i:=h_i\inn L_G-(h_i\inn R^{mn})H_{mn}=\heps_k E^k{}_i\,,           \nn\\
&&E^k{}_i:=2a_0 G^k{}_i=2a_0\Big(R^k{}_i-\frac{1}{2}\d^k_iR\Big)\,.
\eea
Then, the gravitational field equation \eq{2.4c} can be written in an equivalent tensorial form as
\be
2a_0G^k{}_i=\t^k{}_i+\cT^k{}_i\,.                                 \lab{A.4}
\ee
Explicit calculation confirms the validity of Eq. \eq{A.4}.


\end{document}